\documentclass[journal=cmatex,manuscript=article]{achemso}

\usepackage{chemformula} 
\usepackage[T1]{fontenc} 

\author{Erik Pollmann}
\email{erik.pollmann@uni-due.de}
\author{Lukas Madau\ss}
\author{Simon Schumacher}
\altaffiliation{present affiliation: Technical Chemistry III – Faculty of Chemistry, University of Duisburg-Essen, Carl-Benz-Straße 199, 47057 Duisburg, Germany}
\author{Uttam Kumar}
\altaffiliation{School of Material Science and Engineering, University of New South Wales, AUS-2052 Sydney, Australia}
\author{Flemming Heuvel}
\author{Christina vom Ende}
\author{S\"{u}meyra Yilmaz}
\author{S\"{u}meyra G\"{u}nd\"{o}rm\"{u}s}
\author{Marika Schleberger}
\affiliation{Faculty of Physics and CENIDE, University of Duisburg-Essen, Lotharstra\ss e 1, 47057 Duisburg, Germany}

\title[Apparent Differences between Single Layer Molybdenum Disulfide Fabricated via Chemical Vapor Deposition and Exfoliation]
  {Apparent Differences between Single Layer Molybdenum Disulfide Fabricated via Chemical Vapor Deposition and Exfoliation}

\abbreviations{2D, MoS$_{2}$, MoO$_{3}$, AHM, TMDC, CVD, vdW, PL, AFM}

\begin{document}


\begin{abstract}
Innovative applications based on two-dimensional solids require cost-effective fabrication processes resulting in large areas of high quality materials. Chemical vapour deposition is among the most promising methods to fulfill these requirements. However, for 2D materials prepared in this way it is generally assumed that they are of inferior quality in comparison to the exfoliated 2D materials commonly used in basic research. In this work we challenge this assumption and aim to quantify the differences in quality for the prototypical transition metal dichalcogenide MoS$_2$. To this end single layers of MoS$_{2}$ prepared by different techniques (exfoliation, grown by different chemical vapor deposition methods, transfer techniques, and as vertical heterostructure with graphene) are studied by Raman and photoluminescence spectroscopy, complemented by atomic force microscopy. We demonstrate that as-prepared MoS$_{2}$, directly grown on SiO$_{2}$, differs from exfoliated MoS$_{2}$ in terms of higher photoluminescence, lower electron concentration, and increased strain. As soon as a water film is intercalated (e.g., by transfer) underneath the grown MoS$_{2}$, in particular the (opto-)electronic properties become practically identical to those of exfoliated MoS$_{2}$. A comparison of the two most common precursors shows that the growth with MoO$_{3}$ causes greater strain and/or defect density deviations than growth with ammonium heptamolybdate. As part of a heterostructure directly grown MoS$_{2}$ interacts much stronger with the substrate, and in this case an intercalated water film does not lead to the complete decoupling, which is typical for exfoliation or transfer. Our work shows that the supposedly poorer quality of grown 2D transition metal dichalcogenides is indeed a misconception.

\end{abstract}

\section{Introduction}

Two-dimensional (2D) materials are in general thin, flexible, and transparent and it is expected that the semiconducting ones among them will soon play an important role in the development of flexible (opto)electronics. In particular, the possibility to stack different 2D materials to design artificial solids with tailored and otherwise inaccessible (opto)electronic properties, has opened the door for a technological breakthrough. However, one of the biggest obstacles that must be overcome before any dreams of real-world application may come true is the manufacturing process: While researchers rely heavily on isolating techniques yielding individual crystallites of 2D materials, industrial implementation requires growth techniques which are scalable, cost-effective, and reliable.

Several hundred 2D materials could be realized so far, but the transition metal dichalcogenides (TMDCs) are the most intensively studied. This is certainly due to their physical properties, such as a large band gap in the visible range, but also to the fact that they are relatively easy to produce. Like graphene, TMDC monolayers can be isolated, e.g., by mechanical exfoliation \cite{Novoselov.2004, Novoselov.2005} and can be grown, e.g., via chemical vapour deposition (CVD) \cite{Lee.2012, Dumcenco.2015, Han.2015}. Like in the case of graphene, most basic research is still done with exfoliated flakes as it is commonly believed that their quality exceeds that of CVD grown flakes. However, the term \textit{quality} usually refers to defect densities, and it has already been shown that 2D materials grown by CVD can exhibit defect densities which are comparable to those of exfoliated material \cite{Hong.2015}. Thus, for graphene, being metallic, the common belief could indeed be true, as any crystalline defects are detrimental for its outstanding transport properties. For the optoelectronic properties of a semiconducting 2D material, however, appropriate quality criteria might be quite different. For example, the comparison between CVD grown MoS$_{2}$ with exfoliated MoS$_{2}$, both on sapphire \cite{Dumcenco.2015}, or in this study on SiO$_{2}$ (see Fig. \ref{Fig:Samples} below), shows a much higher photoluminescence (PL) intensity in the case of CVD growth. Not as obvious as the intensity of a PL signal, but equally important, are the changes of the Raman mode positions for MoS$_{2}$ fabricated from CVD and exfoliation, which can be directly attributed to physical properties such as strain and doping \cite{Rice.2013, Wang.2013, Conley.2013, Castellanos-Gomez.2013, Scheuschner.2014, Lloyd.2016, Kukucska.2017, Christopher.2019, Chakraborty.2012}.

A pertinent problem in this context is that any comparison between 2D materials from CVD and exfoliated 2D materials requires that the materials be measured under the same or at least similar environmental conditions. Otherwise, if distinct differences are revealed, they can either be attributed to the material's quality, e.g., preparation-specific inhomogeneities, but they could also originate from preparation conditions or the chosen substrate \cite{Michail.2016, Chae.2017, Dubey.2017}, as 2D materials are extremely sensitive to their immediate surroundings. For van der Waals (vdW) heterostructures this becomes even more critical as a number of interfaces are involved which may involve any combination of exfoliated and CVD grown materials. The lack of spectroscopic data taken on differently prepared samples under the strict condition of similar environmental conditions renders it currently almost impossible to decide, how much of what has been derived from experiments on exfoliated materials, or whether any of the reported proof-of-principle concepts may be transferred to CVD grown material at all. 

In this paper we fill this crucial gap and clarify the origin of apparent spectroscopic (PL, Raman) differences found for MoS$_{2}$ either grown by CVD or isolated by exfoliation. For our comprehensive study we use exfoliated MoS$_{2}$ as a reference system, grow samples via CVD using two different precursors (molybdenum trioxid (MoO$_{3}$) and ammonium heptamolybdate (AHM)) and apply transfer techniques to obtain similar environmental conditions for MoS$_{2}$ fabricated by exfoliation and by CVD, respectively. Our experimental data shows that as prepared, MoS$_{2}$ directly grown on SiO$_{2}$ differs from exfoliated MoS$_{2}$ in terms of higher photoluminescence, lower electron concentration, and increased strain, seemingly supporting the widely shared belief of supposedly superior quality of exfoliated material. However, adding intercalated water turns the picture upside-down: We provide clear evidence by Raman spectroscopy in conjunction with PL spectroscopy and atomic force microscopy (AFM) that the intrinsic properties of exfoliated and CVD grown are, in fact, almost identical. Any spectroscopically observed differences can unambiguously be attributed to the presence or absence of an intercalated water film, respectively. Our findings are further supported with experiments performed on aged samples and extended to a prototypical MoS$_{2}$-graphene heterostructure. From our results we conclude that MoS$_{2}$ grown with the right precursor (AHM) may serve as a more-than-adequate substitute for exfoliated material.

\section{Results and Discussion}

\begin{figure}[h!tb]
\centering
\includegraphics[width=1.0\textwidth]{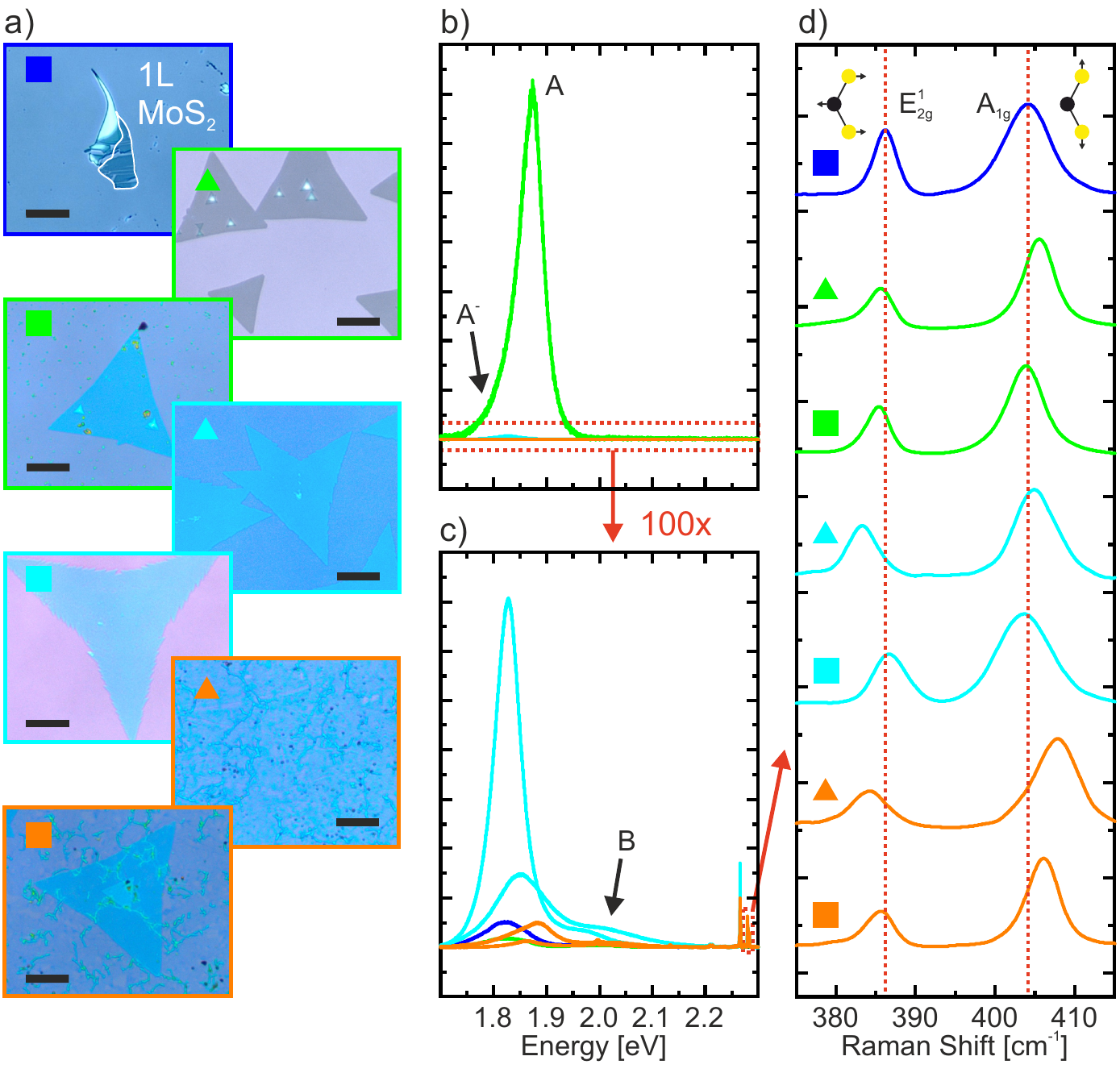}
\caption{Overview of all sample systems. (a) Images of optical microscopy. The black bar is equivalent to 20~\textmu m in all images. Symbols and colors group the samples. Squares: Exfoliated or transferred MoS$_{2}$. Triangles: MoS$_{2}$ grown directly by CVD. Blue: exfoliation. Green: CVD with MoO$_{3}$. Light blue: CVD with AHM. Orange: graphene-MoS$_{2}$ heterostructures. (b) and (c) Representative PL spectra of all samples. (d) E$^{1}_{2g}$ and A$_{1g}$ mode of all samples. Red line indicates mode positions for exfoliated MoS$_{2}$.}
\label{Fig:Samples}
\end{figure}

\paragraph{Overview and Key Question}
We start by introducing the sample systems studied in this paper. Figure~\ref{Fig:Samples} (a) presents the seven different single layer MoS$_{2}$ systems under investigation: exfoliated MoS$_{2}$ on SiO$_{2}$ (blue square), MoS$_{2}$ grown by CVD with AHM precursor on SiO$_{2}$ (light blue triangle) and transferred onto SiO$_{2}$ (light blue square), MoS$_{2}$ grown by CVD with MoO$_{3}$ powder source on SiO$_{2}$ (green triangle) and transferred onto SiO$_{2}$ (green square), MoS$_{2}$ directly grown on transferred CVD graphene (orange triangle), and CVD grown MoS$_{2}$ transferred onto transferred CVD graphene (orange square). The images from the optical microscope are all identical in size  -- the scalebars correspond to 20~\textmu m. This substantiates one of the well-known advantages of CVD over exfoliation as a fabrication method for 2D materials: a significantly larger flake size.

For a first comparison of our sample systems we performed PL and Raman spectroscopy, see Fig.~\ref{Fig:Samples} (b)-(d). The obtained PL spectra in Fig.~\ref{Fig:Samples} (b) and (c) show the characteristic exciton and trion peaks of MoS$_{2}$ \cite{Mak.2010, Splendiani.2010, Mak.2013}. The most prominent peak between 1.8 and 1.9~eV corresponds to the A exciton, the peak around 2.0~eV to the B exciton. The A$^{-}$ trion peak can been found as a shoulder of the A exciton peak around 1.8~eV. Apparently, the spectroscopic signature of single layers of MoS$_{2}$ prepared differently shows a large variation. Note that, while the variation between sample systems is large, the spectra shown in Fig.~\ref{Fig:Samples} are nevertheless typical and thus representative for any given sample system.

\begin{figure}[h!tb]
\centering
\includegraphics[width=1.0\textwidth]{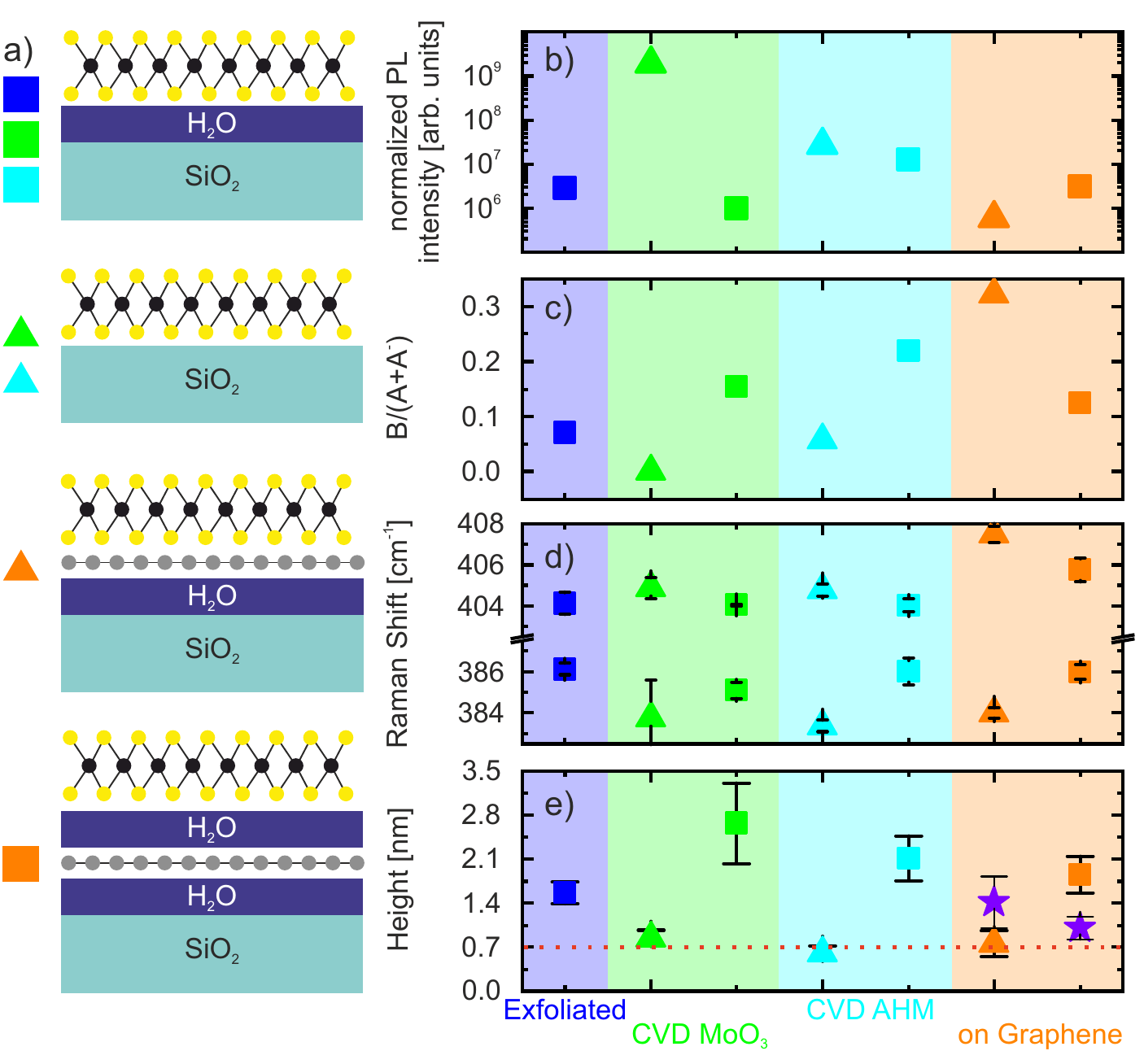}
\caption{Illustration of the sample systems and compilation of all our data. (a) Schematic illustration of the sample systems with an assignment to the real samples by symbols. (b) PL intensities, c) B/(A+A$^{-}$) exciton peak intensity ratio, (d) Raman mode positions, and (e) MoS$_{2}$ (and graphene - purple star) height recorded by AFM for all sample systems.}
\label{Fig:AFM+PL+Raman}
\end{figure}

Because our analysis is based on a comparison of differently prepared samples we present in Fig.~\ref{Fig:AFM+PL+Raman} schematic illustrations of our sample systems together with a compilation of our data before discussing the individual samples further below. In Fig.~\ref{Fig:AFM+PL+Raman} (b) and (c) the PL intensities (total area of all exciton and trion peaks) and the intensity ratio B/(A+A$^{-}$) for the representative PL spectra from Fig.~\ref{Fig:Samples} are shown, respectively. All intensities are normalized to the applied laser power and the integration time, thus to the number of incident photons. The ratio B/(A+A$^{-}$) refers to the (area) intensity ratio of the peaks of the A exciton, the B exciton, and the A$^{-}$ trion, respectively. Figure~\ref{Fig:AFM+PL+Raman} (d) shows the position of the E$^{1}_{2g}$ (around 385~cm$^{-1}$) and the A$_{1g}$ (around 405~cm$^{-1}$) Raman mode as an average of the acquired Raman spectra. In addition to spectroscopy we performed supportive AFM measurements to determine the thickness of the MoS$_{2}$ layers, see SI for a selection of AFM images and line scans and Fig.~\ref{Fig:AFM+PL+Raman} (e) for the resulting thicknesses. Note that the intercalated water layers depicted in Fig. \ref{Fig:AFM+PL+Raman} (a) were not added on purpose to our sample system. They are nevertheless present and turn out to be the key factor for some of the properties as will be shown further below.

In order to reveal the origin of the systematic variations we will now discuss the individual features in detail by discussing PL, Raman, and AFM results for each sample system. We start by comparing the two most well known single layer MoS$_{2}$ systems: exfoliated (blue squares) and directly grown CVD MoS$_{2}$ (green and light blue triangles), both on SiO$_{2}$. From Fig. \ref{Fig:AFM+PL+Raman} (b) it can be seen that the total PL intensity of CVD grown MoS$_{2}$ is up to three orders of magnitude larger than for the exfoliated counterpart. Therefore, from the PL point of view CVD MoS$_{2}$ is clearly superior to exfoliated MoS$_{2}$. The higher intensities make CVD MoS$_{2}$ the material of choice for optoelectronic applications.

From the PL spectra more information about the MoS$_{2}$ can be extracted. Mak et al. reported a decreasing A/A$^{-}$ ratio with increasing charge carrier concentration in single layer MoS$_{2}$ \cite{Mak.2013}. Actually, the A exciton intensity is almost exclusively influenced by the charge carrier concentration. Although not discussed by Mak et al., but shown by their data, the B exciton as well as the A$^{-}$ trion intensity remains relatively constant with changing charge carrier concentration. As it can be difficult to seperate the A and the A$^{-}$ peak correctly -- they are only a few 10~meV separated from each other -- we use the B/(A+A$^{-}$) ratio instead. From this, it can be qualitatively concluded that a high B/(A+A$^{-}$) ratio means that the system is n-doped. For our comparison of CVD grown MoS$_{2}$ (green and light blue triangles) with exfoliated MoS$_{2}$ (blue square) this means that the latter is more n-doped by charge transfer from the substrate, see Fig.~\ref{Fig:AFM+PL+Raman} (c). Unfortunately, the exciton peak ratio is not linearly dependent on the carrier concentration and the peaks -- both, shifts and intensity ratio change -- can be affected due to several other reasons such as defects \cite{Tongay.2013, Nan.2014}, strain \cite{Conley.2013, Castellanos-Gomez.2013, Lloyd.2016, Frisenda.2017, Steinhoff.2017, Christopher.2019}, incident laser power \cite{Tongay.2013, Kaplan.2016, Chen.2017, Steinhoff.2017}, and dielectric screening \cite{Lin.2014, Steinhoff.2017}. Thus, the PL spectra are not suitable for a quantitative analysis, but will be used supportively in the following.

Next, we will show that the commonly used rule for single layer characterization by Raman spectroscopy has to be adapted for CVD material. The typical approach to interpret Raman spectra of MoS$_{2}$ is to determine the difference between the position of the E$^{1}_{2g}$ and the A$_{1g}$ mode. The result gives an information about the number of layers: for exfoliated MoS$_{2}$ a difference of $\sim$19~cm$^{-1}$ corresponds to single layer, while a difference of $\sim$22~cm$^{-1}$ is already attributed to a bilayer \cite{Lee.2010}. Here, we find a difference of 18.0~$\pm$~0.75~cm$^{-1}$ for exfoliated MoS$_{2}$ (blue square), 21.1~$\pm$~1.55~cm$^{-1}$ (green triangle), and 21.4~$\pm$~0.46~cm$^{-1}$ (light blue triangle) for the MoS$_{2}$ grown by the two different CVD methods, see Fig. \ref{Fig:AFM+PL+Raman} (d). The latter wave numbers are already quite close to the number for a bilayer in the case of exfoliated samples. However, the high PL intensity is a clear indication of the single layer nature of the grown material \cite{Splendiani.2010, Mak.2010, Kuc.2011}. This is further corroborated by the AFM data from Fig. \ref{Fig:AFM+PL+Raman} (e). It shows that the determined thickness of the directly grown MoS$_{2}$ flakes (triangles) corresponds well with the expected thickness of one layer of MoS$_{2}$ ($\sim$~0.7~nm). This is clear evidence, that the CVD MoS$_{2}$ is indeed single layer. Exfoliated MoS$_{2}$ flakes (blue square) are frequently found to be much thicker, but are identified as single layer from the Raman data. The reason for the increased thickness is most likely due to intercalated water between substrate and 2D material, which is often reported for exfoliated 2D materials \cite{Novoselov.2005, Xu.2010, Komurasaki.2012, Ochedowski.2013, Ochedowski.2014, Temmen.2014, Varghese.2015}. In contrast, during the CVD process the presence of a water film appears to be highly unlikely because temperatures up to 800~\textdegree C are present. As a result, the grown 2D material is in direct contact with the substrate and its true height might be accessed via AFM. The reason for the unusually large difference between the two Raman peaks in CVD grown MoS$_{2}$ is thus not due to the presence of bilayer. Rather, the Raman mode positions are also affected by strain \cite{Rice.2013, Wang.2013, Conley.2013, Castellanos-Gomez.2013, Scheuschner.2014, Lloyd.2016, Kukucska.2017, Christopher.2019} and the charge carrier concentration \cite{Chakraborty.2012, Kukucska.2017}. The shift of the E$^{1}_{2g}$ mode to higher and of the A$_{1g}$ mode to lower values, respectively, indicate qualitatively that the CVD material is exposed to higher strain and exhibits a lower charge carrier concentration. This will be discussed in more detail below.

So far we have seen that exfoliated and CVD grown MoS$_{2}$ obviously differ in PL intensity, but also in doping and strain. This leads us to one of the key question of this paper. It is usually assumed that 2D materials grown by CVD directly as single layer have a lower structural quality than exfoliated materials, which are claimed to have a remarkably high quality \cite{Novoselov.2004, Novoselov.2005}. The defect density and the grain size will of course affect the material's properties but is this, and therefore the production method, really the reason for the observed differences in the properties of CVD grown MoS$_{2}$ and exfoliated MoS$_{2}$?

In order to answer this question we prepare sample systems from CVD grown MoS$_{2}$ which can be directly compared to exfoliated MoS$_{2}$. To this end CVD MoS$_{2}$ flakes were transferred from their original growth substrate SiO$_{2}$ onto a second, clean SiO$_{2}$ substrate (green and light blue squares). The analysis of this system shows a decreased PL intensity, an increased B/(A+A$^{-}$) ratio, a decreased Raman mode difference, and an increased flake thickness, see Fig.~\ref{Fig:AFM+PL+Raman} (b)-(e). This means, that after the transfer of the CVD grown flake, the properties of the material appear now very similar, even almost identical, to those of exfoliated MoS$_{2}$ (for this reason this type of samples is represented by a square in all diagrams, just like exfoliated MoS$_{2}$). This finding is even more intriguing as one would have expected the transfer to rather degrade the overall quality of the material. Based on our results, we can at this point thus formulate two hypotheses: (i) the fabrication method itself has only a very minor influence on the intrinsic properties of single layer MoS$_{2}$, and (ii) the dominant factor for the actual properties of 2D MoS$_{2}$ is the presence or absence of an intercalated water film - which is a purely external effect.

\begin{figure}[h!tb]
\centering
\includegraphics[width=1.0\textwidth]{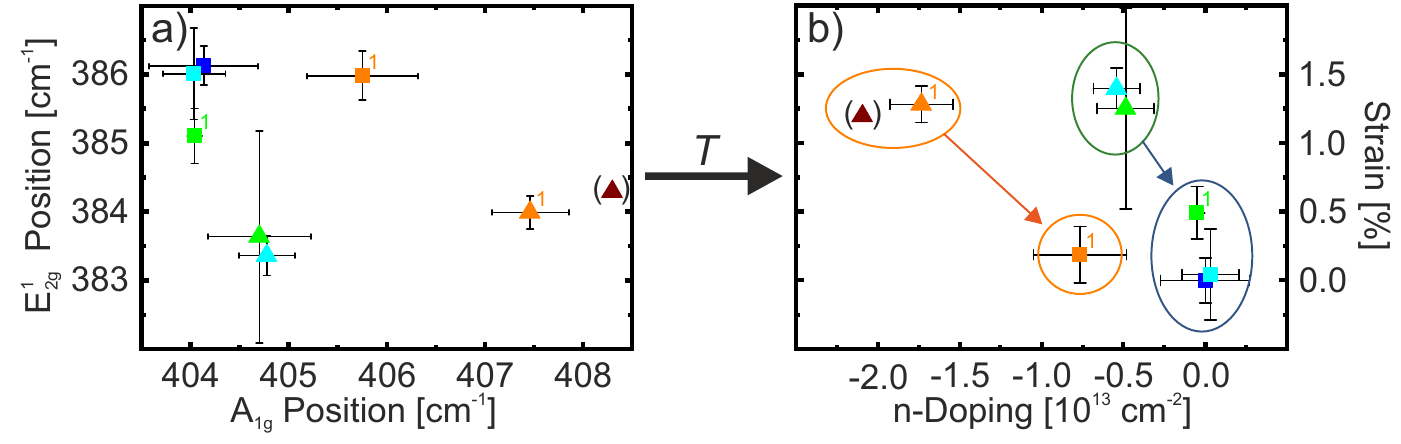}
\caption{Transformation of the diagram of (a) the measured Raman mode positions to (b) the corresponding physical properties (strain, doping). Groups of sample systems with similar properties  become apparent. Brown triangle: MoS$_{2}$ grown on highly oriented pyrolytic graphite \cite{Pollmann.2019}.}
\label{Fig:Transformation}
\end{figure}

\paragraph{Quantification by Transformation Matrix}
Next, we will quantify the physical properties to validate our hypotheses. To this end, we present the Raman mode positions (from Fig. \ref{Fig:AFM+PL+Raman}) for the differently prepared MoS$_{2}$ systems in a different way: In Fig.~\ref{Fig:Transformation} (a) the E$^{1}_{2g}$ mode position is plotted vs. the A$_{1g}$ mode position. This diagram makes it easier to identify groups of sample systems with similar behaviour. For the analysis we construct a matrix $T$ based on literature data taking into account the mode dependency on strain \cite{Rice.2013, Wang.2013, Conley.2013, Castellanos-Gomez.2013, Scheuschner.2014, Lloyd.2016, Kukucska.2017, Christopher.2019} and doping \cite{Chakraborty.2012, Kukucska.2017} allowing us to conduct an axis transformation as given by equation \ref{Eq:Transformation}. All publications that have studied the evolution of Raman mode positions as a function of strain report that strain causes a shift of the E$^{1}_{2g}$, while the A$_{1g}$ mode position is almost exclusively influenced by doping. Biaxial \cite{Scheuschner.2014, Lloyd.2016, Kukucska.2017} and uniaxial strain \cite{Rice.2013, Wang.2013, Conley.2013, Castellanos-Gomez.2013, Kukucska.2017, Christopher.2019}, however, differ with respect to the exact number of wave number shift per percentage of strain. Although MoS$_{2}$ typically deforms in uniaxially strained wrinkels (orders of magnitude larger than the substrate roughness) \cite{Castellanos-Gomez.2013, Lin.2015, Deng.2017, Pollmann.2018, Deng.2019}, we assume that on the nanoscale the deformation of a MoS$_{2}$ layer on a given substrate is a mixture of strained, unstrained, uniaxially and biaxially strained regions averaged by the \textmu m$^{2}$ laser spot. For the construction of the matrix $T$ we use thus the publication of Rice et al. \cite{Rice.2013} and Chakraborty et al. \cite{Chakraborty.2012}, who have quantified the strain- and doping-dependent Raman shift experimentally and by calculations. Note that Rice et al. only report uniaxial strain, which may cause some deviation in the absolute value.

\begin{equation}
\label{Eq:Transformation}
	\begin{pmatrix}
		-0.490 \ \mathrm{\%/cm^{-1}} & -0.073 \ \mathrm{\%/cm^{-1}} \\
		0.088 \times 10^{13} \ \mathrm{cm^{-2}/cm^{-1}} & -0.464 \times 10^{13} \ \mathrm{cm^{-2}/cm^{-1}}
	\end{pmatrix}
	\begin{pmatrix}
		\Delta E^{1}_{2g} \\
		\Delta A_{1g}
	\end{pmatrix}
	=
	\begin{pmatrix}
		\Delta Strain \\
		\Delta Doping
	\end{pmatrix}
\end{equation}

This matrix can now be used to transform our E$^{1}_{2g}$ vs. A$_{1g}$ plot (or mode positions of a single spectrum with respect to a given reference spectrum) into a strain vs. n-doping plot, see Fig.~\ref{Fig:Transformation} (b). Because mode position changes are considered with regard to a reference system, only relative strain and doping are shown. The ideal reference system would be a single layer MoS$_{2}$ system with no strain and the intrinsic doping level. Unfortunately, especially the latter has not been achieved yet. For our discussion here we use exfoliated MoS$_{2}$ as a reference (i.e. zero point for the axes in Fig. \ref{Fig:Transformation} (b)), a choice which can be justified by the following reasons: Exfoliation of single layer MoS$_{2}$ is based on the detachment of the last layer from a bulk crystal. Even if the scotch tape would induce any strain in bulk MoS$_{2}$, the multi layered nature of MoS$_{2}$ in combination with the weak van der Waals forces in between layers ensure that any strain built into the last layer is released upon exfoliation.

From the resulting diagram (Fig.~\ref{Fig:Transformation} (b)) the strain and the n-doping can now be quantified with respect to exfoliated MoS$_{2}$. As stated in our hypothesis and shown  qualitatively before, transferred CVD grown MoS$_{2}$ (green and light blue square) has very similar properties to exfoliated MoS$_{2}$ (blue square) indeed. However, directly grown MoS$_{2}$ (green and light blue triangle) is more strained by 1--1.5~\% and has a reduced electron concentration by approximately 0.55~$\times$~10$^{13}$~cm$^{-2}$. Both property changes can be explained by the absence of an intercalated water layer in the case of directly grown MoS$_{2}$ and will be discussed separately in the following.

\paragraph{Doping}
Focusing on the x-axis in Fig. \ref{Fig:Transformation} (b) we can confirm our hypotheses based on the qualitative analysis (see above) for the property \textit{doping}. First, we find that the intercalated water film between MoS$_{2}$ and SiO$_{2}$ either acts as a donor or efficiently screens the acceptors of the substrate, i.e., it has an n-doping influence on MoS$_{2}$. This is consistent with previous publications which demonstrate by PL that exfoliated MoS$_{2}$ on SiO$_{2}$ is more n-doped compared to MoS$_{2}$ on other substrates \cite{Buscema.2014} or freestanding MoS$_{2}$ \cite{Scheuschner.2014}. In case of MoS$_{2}$ on mica Varghese et al. showed, that the intercalated water film plays a crucial role for the doping level \cite{Varghese.2015}. In addition to comparable PL measurements they performed Kelvin probe microscopy revealing an increased work function of MoS$_{2}$ with an intercalated water film in comparison to MoS$_{2}$ in direct contact with the substrate.

Even more remarkable is how closely the data points in Fig. \ref{Fig:Transformation} (b) match with respect to the average doping level. Transferred CVD MoS$_{2}$ (green and light blue square) deviates from exfoliated MoS$_{2}$ (blue square) by only 5~$\times$~10$^{11}$ charge carriers per cm$^{2}$. The electron concentration of MoS$_{2}$ samples from the different CVD variants (green and light blue triangle) differ by less than 6~$\times$~10$^{11}$~cm$^{-2}$. This analysis clearly shows that neither the CVD process (defects or impurity atoms) nor the transfer (PMMA residues) causes an effective doping relative to exfoliated MoS$_{2}$. As a consequence, we conclude that the various fabrication methods do not influence the intrinsic carrier concentration in MoS$_{2}$.

\begin{figure}[H]
\centering
\includegraphics[width=1.0\textwidth]{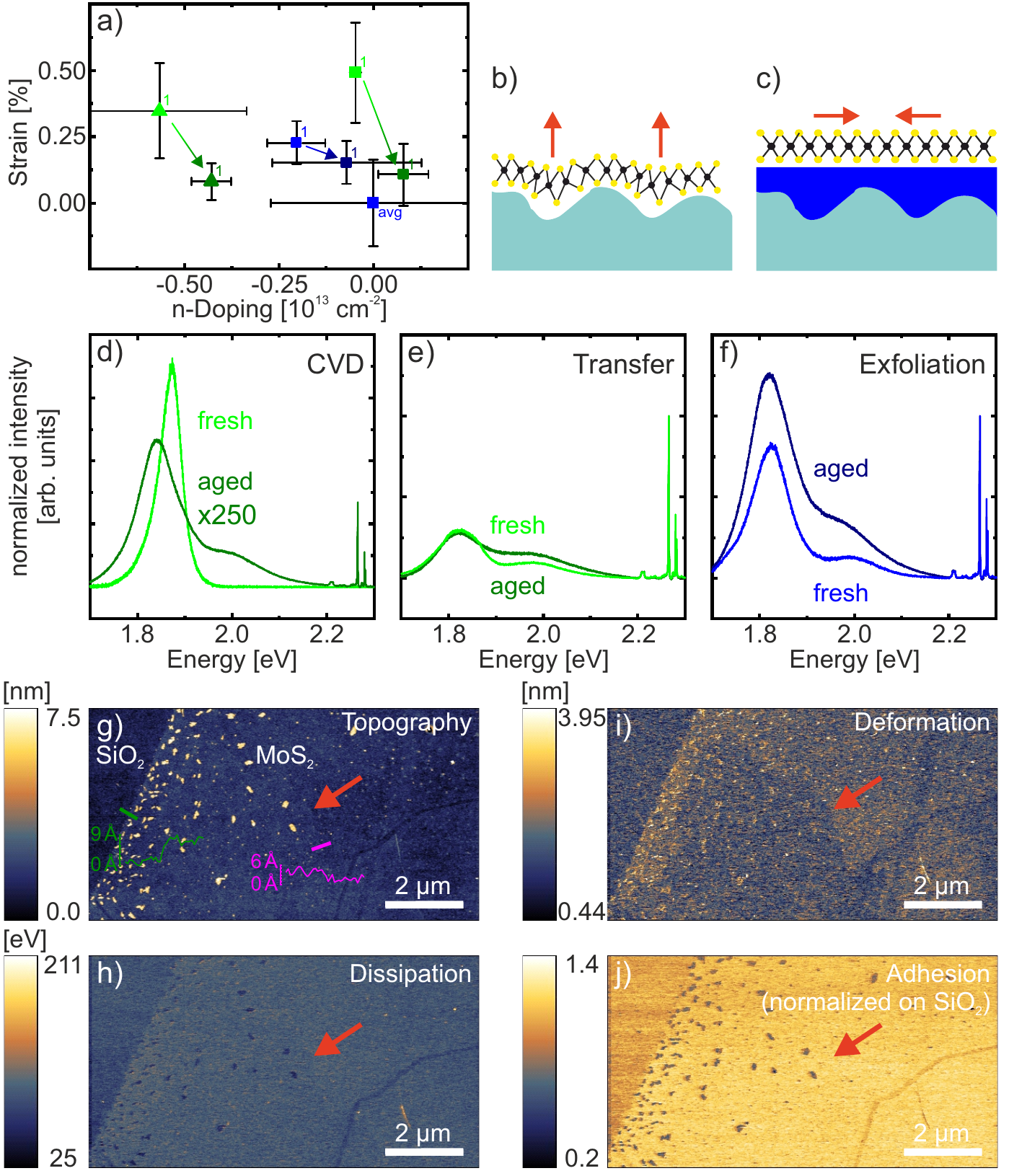}
\caption{Aging effect of differently prepared MoS$_{2}$ samples. (a) Strain and doping values from the Raman mode position before (bright) and after (darker) storage time. Schematic of net forces assumed to be present due to in-plane strain in CVD MoS$_{2}$ (b) as grown and (c) subsequently transferred on SiO$_{2}$. (d)-(f) PL Peaks before (brighter curves) and after (darker curves) aging. PeakForce AFM channels of an aged, directly CVD grown MoS$_{2}$ flake, i.e. (g) topography, (h) dissipation, (i) deformation, (j) adhesion.}
\label{Fig:Aging}
\end{figure}

\paragraph{Strain}
In the following paragraphs we want to discuss the relative \textit{strain} plotted on the y-axis in Fig. \ref{Fig:Transformation} (b). In contrast to directly grown MoS$_{2}$ (green and light blue triangle), for exfoliated (blue square) or transferred MoS$_{2}$ (green and light blue square), respectively, we find lower strain values. One possible explanation for these findings is related to the properties of the substrate. The arithmetic average roughness $R_{a} = 0.2$~nm measured on SiO$_{2}$ is rather high with respect to the MoS$_{2}$ layer thickness (unchanged before and after the MoS$_{2}$ growth process -- see SI). This may lead to a much stronger bending of 2D material sheets in direct contact with the substrate in comparison to 2D material sheets decoupled from the substrate by an intercalated water film. An illustration of this effect is given in Fig. \ref{Fig:Aging} (b) and (c). Both hypotheses are thus confirmed again, but not as strictly as in the case of doping. This has to do with the comparably large variation of strain values, see the large error bar for the green triangle and the large scatter of the squares in Fig. \ref{Fig:Transformation}.

The large variation of the strain value is only observed when comparing MoS$_{2}$ flakes from different batches grown with the MoO$_{3}$ precursor (green triangle), and even if all process parameters are kept constant. However, MoS$_{2}$ flakes from the same batch have very similar Raman mode positions and thus strain values, see Tab.~\ref{Tab:MoO3-Batches}. The large variation of strain values between batches correlates with the poor reproducibility of this growth variant, which may sometimes not even yield single layer MoS$_{2}$, and might be related to the poor control of the exact amount and the relative position of the MoO$_{3}$ powder with respect to the substrate. In contrast, the growth process with AHM as precursor (light blue triangle) is in general much more reproducible and always results in single layer MoS$_{2}$ flakes with rather constant Raman mode positions when comparing different batches. Obviously, the CVD growth does have an influence on the strain dependent E$^{1}_{2g}$ Raman mode of MoS$_{2}$.

\begin{table}
\caption{Raman spectroscopy data from three selected batches of triangular shaped, single layer MoS$_{2}$ successfully grown with MoO$_{3}$ precursor.}
\begin{tabular}{lllll}
\textbf{Batch \#} & \textbf{E$^{1}_{2g}$ [cm$^{-1}$]} & \textbf{A$_{1g}$ [cm$^{-1}$]} & \textbf{rel. Strain [\%]} & \textbf{rel. n-Doping [10$^{13}$ cm$^{-2}$]} \\
\hline 
1 & $385.58 \pm 0.39$ & $405.25 \pm 0.56$ & $0.35 \pm 0.18$ & $-0.57 \pm 0.23$ \\  
2 & $381.89 \pm 0.29$ & $404.48 \pm 0.05$ & $2.10 \pm 0.10$ & $-0.53 \pm 0.04$ \\  
3 & $383.47 \pm 0.18$ & $404.38 \pm 0.25$ & $1.31 \pm 0.09$ & $-0.36 \pm 0.11$ \\
all & $383.64 \pm 1.54$	& $404.70 \pm 0.53$ & $1.25 \pm 0.73$ & $-0.49 \pm 0.17$ \\
\end{tabular}
\label{Tab:MoO3-Batches}
\end{table}

Another cause that exclusively shifts the E$^{1}_{2g}$ mode are sulfur vacancies \cite{Parkin.2016}. Because of the very similar change of the Raman signature of MoS$_{2}$ it is reasonable that strain and sulfur vacancies may be related: the missing atoms cause the other lattice atoms to reorient and thus lead to local strain in the lattice. In our previous study we have shown that MoS$_{2}$ grown by CVD is indeed MoS$_{2-x}$ \cite{Madau.2018}, i.e. the major defect type are sulfur vacancies, which has also been reported by another comparative study \cite{Hong.2015}. Therefore, it is very likely that the large error bar of the E$^{1}_{2g}$ mode, and the corresponding strain value, respectively,  are due to varying sulfur vacancy densities in different MoS$_{2}$ samples grown with the MoO$_{3}$ precursor, i.e. due to different stoichiometries of MoS$_{2-x}$.

Because different MoS$_{2}$ flakes from one batch of the MoO$_{3}$ process variant (green triangle) have similar Raman spectra, but can strongly differ from MoS$_{2}$ flakes of another batch (Tab.~\ref{Tab:MoO3-Batches}), it is thus reasonable that the stoichiometry changes due to, e.g., slightly different (and uncontrollable) source material concentrations during the growth process. In the extreme case the process environment is in a state too far away from a stable stoichiometry of MoS$_{2-x}$, so that no growth takes place or the resulting MoS$_{2-x}$ immediately degrades again. Otherwise, different sulfur vacancy concentrations are possible, thus different E$^{1}_{2g}$ mode positions are found. Because the E$^{1}_{2g}$ mode position may not only depend on the strain but also on the defect density, possibly associated with the strain, we can estimate the maximum variation of defect density of both process variants according to the experimental data of Parkin et al. \cite{Parkin.2016}. We find for MoO$_{3}$ precursor based CVD MoS$_{2}$ (green triangles) a deviation of sulfur vacancies of about $\pm$~0.79~\%, while for MoS$_{2}$ from the AHM process (light blue) a value of only $\pm$~0.15~\% is determined. It is quite remarkable, that these different defect concentrations do not seem to have any effect on the doping. This is however in agreement with DFT calculations showing that the sulfur defect states are located to deep in the band gap to cause doping \cite{Komsa.2015}.

Note, that transfer can be used for a reduction of strain, see green and light blue squares in Fig. \ref{Fig:Transformation} (b). Our data shows that strain in CVD grown MoS$_{2}$ sheets due to a rough SiO$_{2}$ substrate is however only partially released when the substrate is removed, see the deviation of the green square (one sample) with respect to the blue and light blue squares and the large error bar of the light blue square in Fig. \ref{Fig:Transformation} (b). The polymer -- which can introduce additional strain into the MoS$_{2}$ by forming bubbles and wrinkles within the polymer layer \cite{Lin.2015} -- obviously maintains the in-built strain even if the 2D material is transferred to a new substrate with an intercalated water layer, see schematic in Fig. \ref{Fig:Aging} (c).

In summary, we can basically confirm both hypotheses: we found that the fabrication method has no particular influence on the intrinsic doping level of MoS$_{2}$. In contrast, the sulfur vacancy density in CVD MoS$_{2}$, which can be monitored by the strain sensitive E$^{1}_{2g}$ Raman mode, depends on the growth conditions during the process. Nevertheless, the major factor for the apparent difference between CVD grown and exfoliated MoS$_{2}$ is \textit{extrinsic}: intercalated water. The following paragraphs provide further insight into the role of intercalated water and  confirms that exfoliated MoS$_{2}$ is indeed a good reference because there is only little strain, if any.

\paragraph{Aging} We start this additional discussion with some observations from a detailed (although non-comprehensive) aging study (Fig.~\ref{Fig:Aging}). It For this study a CVD grown MoS$_{2}$ sample with a low defect density (green triangles), a transferred CVD MoS$_{2}$ sample (green squares), and an exfoliated MoS$_{2}$ sample (blue squares) were stored for several months under inert nitrogen atmosphere at a humidity of about 20~\% to avoid degradation \cite{Gao.2016}. We found that particularly the strain in both CVD based systems was reduced to the level of exfoliated MoS$_{2}$, while the strain of the exfoliated reference sample remained constant (Fig. \ref{Fig:Aging} (a)). The altered strain in the transferred CVD MoS$_{2}$ sample towards the strain level of exfoliated MoS$_{2}$ confirms the hypothesis of strain being introduced by the polymer during the transfer.

A possible explanation is illustrated in Fig. \ref{Fig:Aging}~(b) and (c): In transferred MoS$_{2}$ (Fig. \ref{Fig:Aging} (c)) only in-plane forces are exerted by strain, which let the MoS$_{2}$ layer relax on its buffer water layer. In contrast, the net forces in CVD MoS$_{2}$ can be more complex because of the rough substrate (Fig. \ref{Fig:Aging} (b)). When the MoS$_{2}$ relaxes, channels may form between the MoS$_{2}$ layer and the substrate, which would be partially filled up with water from the residual humidity. Two additional observations were made in our aging study: (i) The doping level for all sample systems changes only marginally (Fig.~\ref{Fig:Aging}~(a)), (ii) the PL intensity only changes significantely for CVD MoS$_{2}$ (green triangles), see Fig. \ref{Fig:Aging} (d)-(f), hence with respect to optoelectronic properties this sample system is most sensitive to storage time. Note: Raman and PL spectra were obtained at different locations and on different flakes of the stored samples, but obviously no spatial variations were detected.

\paragraph{Intercalated Water}
To test the hypothesis of the strain reduction mechanism in aged CVD MoS$_{2}$ we performed AFM measurements shown in Fig. \ref{Fig:Aging} (g)-(j). Here we take advantage of the so-called PeakForce Tapping mode, which allows us to  additionally obtain nanomechanical properties such as \textit{deformation} (Fig. \ref{Fig:Aging} (i)) and \textit{adhesion} (Fig. \ref{Fig:Aging} (j)). Already when looking at the more common channels \textit{topography} (Fig. \ref{Fig:Aging} (g)) and \textit{dissipation} (Fig. \ref{Fig:Aging} (h)) the effect of aging can be clearly seen (red arrow pointing on the same spot in all images). While some parts of the flake are still found to be 0.7~nm thick (green line profile), most areas show a slight increase in height by 2-4~\AA \ (magenta averaged line profile) and exhibit a reduced dissipation (corresponding to a reduction of inelastic deformation by the AFM tip). The nanomechanical data confirms this further because the deformation of MoS$_2$ decreases in those areas. The adhesion force between AFM tip and the MoS$_{2}$ surface, in contrast, remains constant over the whole sample area. At this point we can exclude the possibility that the contrast changes of the topography, dissipation, and deformation channels are caused by an adsorbate layer on top of the MoS$_{2}$ layer because adhesion, which is a measure for the force needed to remove the tip from the surface, remains unchanged. Instead, the slightly higher regions are areas of MoS$_{2}$, underneath which a film of intercalated water already exists. The increase in height by a few Angstroms is in good agreement with the characteristic thickness of a single water molecule layer underneath 2D materials, which was determined to be 0.37~nm \cite{Xu.2010, Komurasaki.2012}. The reduced deformation by the AFM tip in this region shows that this water film is incompressible.

\begin{figure}[h!tb]
\centering
\includegraphics[width=1.0\textwidth]{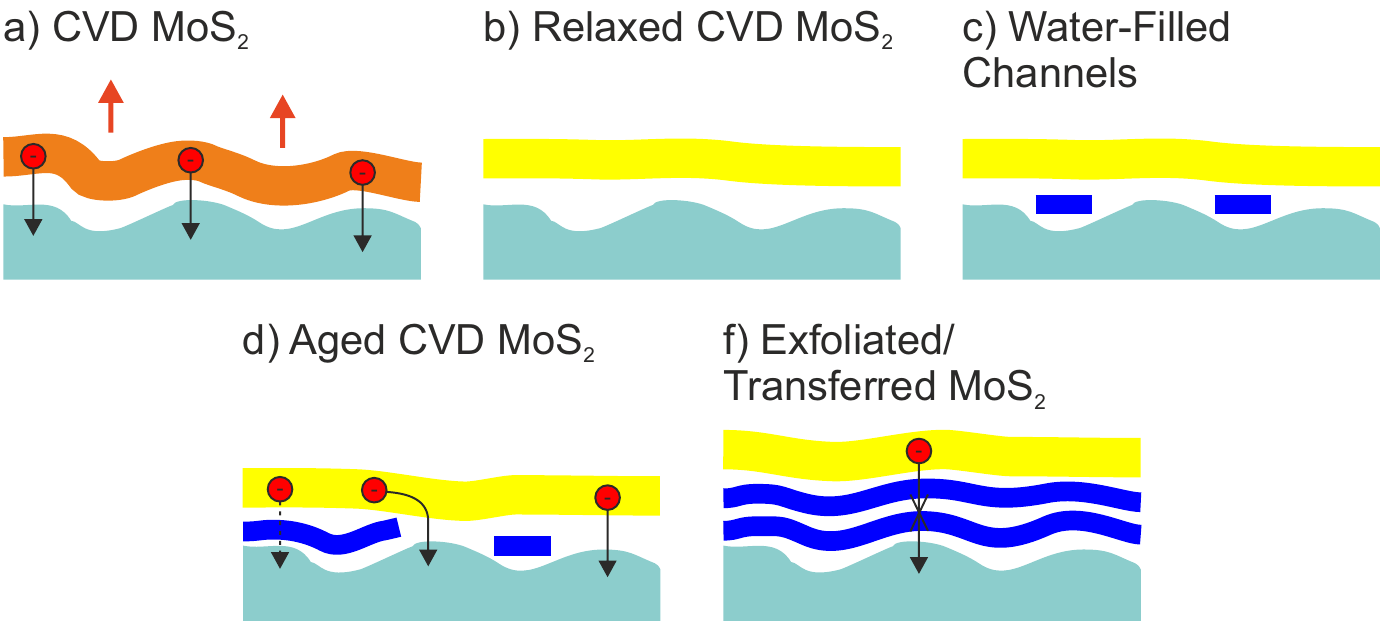}
\caption{Proposed water intercalation mechanism. (a) Single layer MoS$_{2}$ as grown by CVD. In-plane strain (indicated by orange in contrast to the yellow colouring) results in  out-of-plane net forces. (b) By relaxing the CVD MoS$_{2}$ layer channels between MoS$_{2}$ and the substrate are formed. (c) These channels are filled with water due to capillary forces. (d) The water intercalation expands from the water-filled channels forming a water layer. As the water layer is non-continuous and thin charge transfer between MoS$_{2}$ layer and substrate is still possible. (e) Underneath exfoliated or transferred MoS$_{2}$ two or more water layers are intercalated, thus the charge transfer to the substrate is strongly screened.}
\label{Fig:WaterIntercalation}
\end{figure}

For the water intercalation we propose the following mechanism, which is shown step-by-step in Fig.~\ref{Fig:WaterIntercalation}. As grown CVD MoS$_{2}$ is in direct contact with the rough SiO$_{2}$ substrate, causing in-plane strain (Fig.~\ref{Fig:WaterIntercalation}~(a)). This results in net forces which partially detach the MoS$_{2}$ layer from the SiO$_{2}$ substrate and open small channels (Fig.~\ref{Fig:WaterIntercalation}~(b)). These channels are subsequently filled with water due to capillary forces (Fig.~\ref{Fig:WaterIntercalation}~(c). It is very likely that this filling even supports the channel formation and thus the strain reduction, i.e., step (b) and (c) take place simultaneously. Over time the intercalated water in the channels expands to form water layers underneath the MoS$_{2}$ (Fig.~\ref{Fig:WaterIntercalation}~(d)). The latter stage is observed in the AFM images in Fig.~\ref{Fig:Aging}~(g)-(j).

If, as with the aged CVD MoS$_{2}$ sample (green triangle), the water film is non-continuous and/or very thin, charge transfer between MoS$_{2}$ and substrate may be still present (Fig.~\ref{Fig:WaterIntercalation}~(d)). However, if the intercalated water film is continuous and thicker, as in the exfoliated or transferred MoS$_{2}$-SiO$_{2}$ system (> 2 water layers of $\sim$0.37~nm each, illustration: Fig.~\ref{Fig:WaterIntercalation}~(e), AFM data: Fig.~\ref{Fig:AFM+PL+Raman}~(e)), the charge transfer is more efficiently screened resulting in a different doping level in MoS$_{2}$. A strong PL quenching without a change of the doping level is due to the complex mechanisms which influence the excitons in 2D MoS$_{2}$. The thin water film, which allows charge transfer between MoS$_{2}$ and the substrate, may affect the excitons by, e.g., changing the dielectric environment ($\varepsilon_{H_{2}O} > \varepsilon_{SiO_{2}}$), which increases the dissociation of excitons \cite{Steinhoff.2017}, probably supported by the exciton diffusion length \cite{Wang.2012}. Nonetheless, PL intensity and doping are obviously not always strictly connected.

\paragraph{MoS$_{2}$-Graphene Heterostructures}

\begin{figure}[h!tb]
\centering
\includegraphics{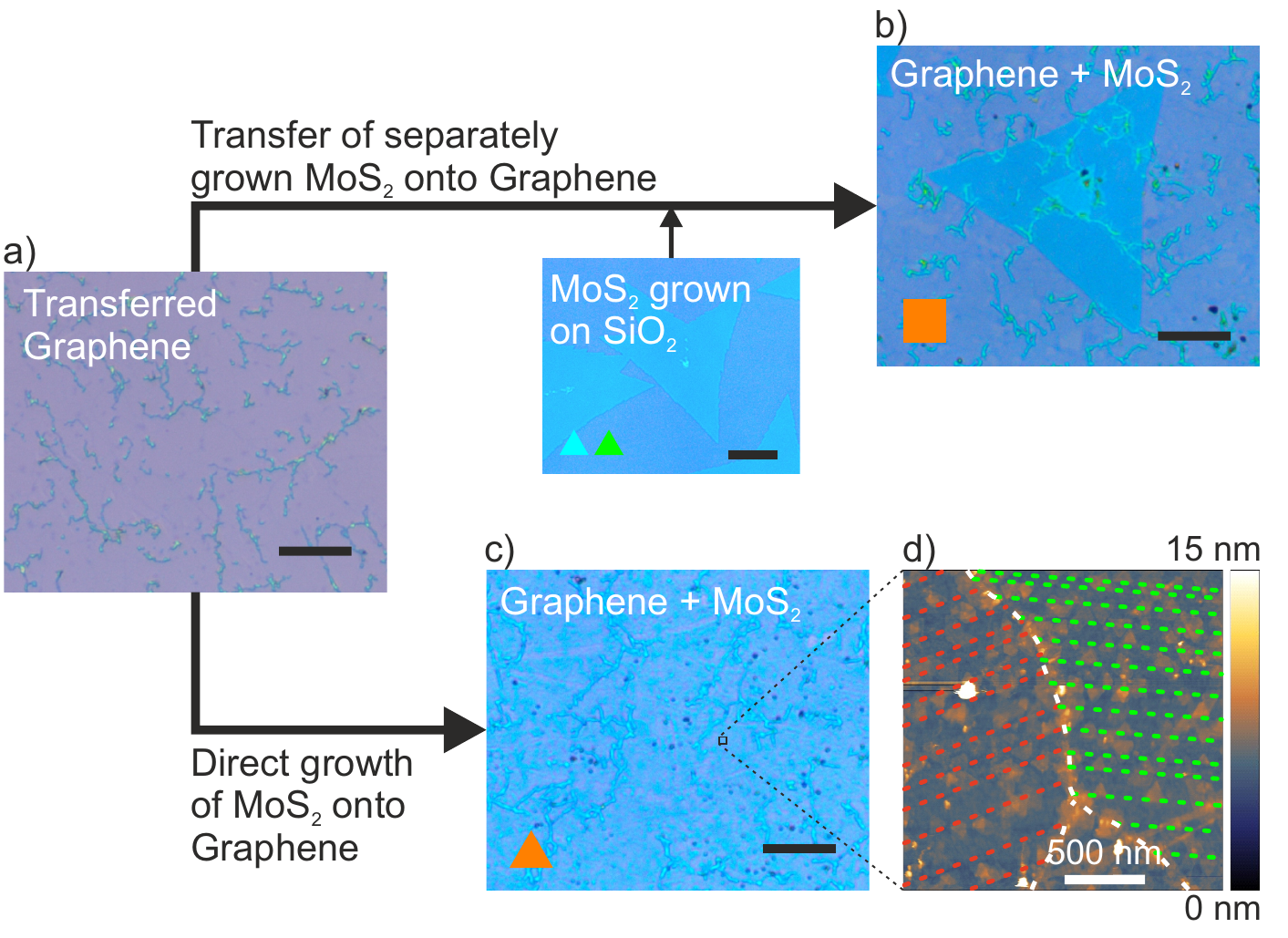}
\caption{Preparation routes of MoS$_{2}$-graphene heterostructures. (a) graphene as transferred onto an SiO$_{2}$ substrate as discribed in our previous study \cite{Madau.2019}. (b) Optical image of resulting MoS$_{2}$-graphene heterostructure by transferring grown MoS$_{2}$ onto graphene. (c) Optical image of MoS$_{2}$-graphene heterostructure resulting from direct growth of MoS$_{2}$ on graphene. (d) AFM (topography) image of triangular MoS$_{2}$ nanoflakes on graphene. Black scale bars correspond to 20~\textmu m.}
\label{Fig:MoS2-Graphene}
\end{figure}

Finally, we have prepared MoS$_{2}$-graphene heterostructures in order to gain further insight to the screening effect of the intercalated water film underneath transferred single layer MoS$_{2}$. The heterostructures have been prepared by two different routes, see Fig.~\ref{Fig:MoS2-Graphene}. Both routes start with CVD graphene (Fig. \ref{Fig:MoS2-Graphene} (a)), which has been transferred from its growth substrate to SiO$_{2}$ by our recently reported polymer-free transfer technique \cite{Madau.2019}. Afterwards, MoS$_{2}$ is either transferred onto (Fig. \ref{Fig:MoS2-Graphene} (b)) or grown by CVD on the resulting graphene-SiO$_{2}$ substrates (Fig. \ref{Fig:MoS2-Graphene} (c) and (d)). In this way, we can compare two different heterostructures: one, where MoS$_{2}$ is in direct contact with graphene and the second one, where a water film with a comparable thickness to exfoliated or transferred MoS$_{2}$ on SiO$_{2}$ is intercalated, see Fig.~\ref{Fig:AFM+PL+Raman}~(a) for illustrations. 

The data and resulting properties can be found in Fig.s \ref{Fig:Samples}-\ref{Fig:Transformation} using orange as color and the same convention for the symbol shape as before, i.e. data points for directly grown MoS$_{2}$ on graphene are represented by the orange triangles and MoS$_{2}$ transferred on graphene is represented by the orange squares. Note that for both heterostructure types the graphene layer has an increased thickness due to intercalated water (Fig. \ref{Fig:AFM+PL+Raman} (e) purple stars), which remains trapped between graphene and SiO$_{2}$ despite high temperatures during the MoS$_{2}$ growth process. This is not surprising, as other studies have shown that water intercalated underneath graphene remains trapped after annealing even under UHV conditions \cite{Ochedowski.2014, Temmen.2014}.

First, we discuss the morpohology of the resulting MoS$_{2}$-graphene heterostructures. For the transferred heterostructure we obtain large areas (up to a few hundreds of \textmu m$^{2}$) of optically perfect heterostructures, see Fig. \ref{Fig:MoS2-Graphene} (b). In the case of directly grown MoS$_{2}$, the whole graphene layer has a more bluish color (Fig. \ref{Fig:MoS2-Graphene} (c)), which is due to MoS$_{2}$ nanoflakes on the graphene surface as revealed by AFM (Fig. \ref{Fig:MoS2-Graphene} (d)).

These triangular single layer MoS$_{2}$ nanoflakes grow in a preferred orientation within a given graphene domain (the boundary is marked  by the white dotted line). At the domain boundaries a particularly large number of MoS$_{2}$  flakes is found. It is typical for MoS$_{2}$ to preferentially grow at imperfections such as step edges as demonstrated for  ``bulk-graphene'' highly oriented pyrolytic graphite (HOPG) substrates \cite{Koos.2016, Lu.2015, Pollmann.2019} and at artificially induced nm-sized defects \cite{Pollmann.2018b}. The oriented growth of MoS$_{2}$ underlines the cleanliness of the graphene surface after our novel transfer technique \cite{Madau.2019}. It is worth noting, that the MoS$_{2}$-graphene system based on directly grown MoS$_{2}$ (orange triangle) has very similar Raman modes in comparison to MoS$_{2}$ grown on HOPG (brown triangles)\cite{Pollmann.2019}. The similarities of MoS$_{2}$ grown on HOPG and on graphene (flake size, nucleation at grain boundaries/step edges, orientation, Raman mode positions) confirm the claim that MoS$_{2}$ on HOPG is a good model system for MoS$_{2}$ on graphene \cite{Koos.2016, Pollmann.2019}.

Now let us turn to the spectroscopic results. The trend is very similar to the MoS$_{2}$-SiO$_{2}$ system, i.e., MoS$_{2}$-graphene prepared by transfer (orange squares) is more n-doped and less strained than MoS$_{2}$-graphene grown directly (orange triangles), see Fig.~\ref{Fig:Transformation}. However, there is an offset to the corresponding MoS$_{2}$-SiO$_{2}$ systems. It is not surprising that MoS$_{2}$ in direct contact with graphene has a different doping level than Mos$_2$ in contact with SiO$_{2}$ because the electronic structures of graphene and SiO$_{2}$ are very different. Because graphene is reported to be p-doped on SiO$_{2}$ under ambient conditions \cite{Novoselov.2004, Ernst.2016}, it seems likely that graphene affects the doping  level MoS$_{2}$ on top towards p-doping. The fact that MoS$_{2}$-graphene heterostructure prepared by transfer is less n-doped than MoS$_{2}$ transferred on SiO$_{2}$ demonstrates that the intercalated water (due to transfer or exfoliation) may not completely screen the substrate. Some charge transfer is still present i.e. the substrate affects the MoS$_{2}$ layer despite the intercalated water.

The influence of the substrate (and the role of the intercalated water) can also be seen by the PL intensity of the differently prepared MoS$_{2}$-graphene heterostructures (Fig.~\ref{Fig:AFM+PL+Raman}~(b)). MoS$_{2}$ transferred onto graphene (orange triangles) exhibits a PL intensity on the same level of MoS$_{2}$ transferred or exfoliated onto SiO$_{2}$ (green, light blue, blue squares), but is doped more like MoS$_{2}$ grown on SiO$_{2}$ (green, light blue triangles), see Fig.~\ref{Fig:Transformation}~(b). Obviously the PL intensity of MoS$_{2}$ is indeed not strictly related to its doping level, but rather to direct environment of the MoS$_{2}$ layer.

This is even more emphasized in the case of MoS$_{2}$ directly grown on graphene (orange triangle). Although it is the most p-doped MoS$_{2}$ system studied in this paper (Fig.~\ref{Fig:Transformation}~(b)) and should thus have an extremely high PL intensity \cite{Mak.2013}, it has the lowest measured PL intensity (Fig.~\ref{Fig:AFM+PL+Raman}~(b)) -- even if we take into account that only a fraction of the graphene area is actually covered with single layer MoS$_{2}$. In this case, the origin of the PL quenching is the dissociation of the excitons by the underlying semimetallic graphene  \cite{Pierucci.2016, Zhang.2014, Huo.2015}. As previously shown for comparable MoSe$_{2}$-graphene heterostructures, the most efficient charge transfer and thus an even stronger PL quenching is to be expected for a direct coupling between the two van der Waals materials \cite{Froehlicher.2018}. Apparently, the direct growth of MoS$_{2}$ on graphene via CVD enables such a direct coupling and once more underlines the potential of this fabrication methods also for high quality van der Waals heterostructures.

\section{Conclusion}

Exfoliated and CVD grown MoS$_{2}$ monolayers appear extremely different -- but this is true only at first glance. The CVD material shows a much more intense PL and the mode positions in its Raman spectra differ significantly from those of the exfoliated material. As a consequence, the usual method for evaluating the number of layers, as described by Lee et al. \cite{Lee.2010} and established on exfoliated samples is not valid for MoS$_{2}$ grown by CVD. From our spectroscopic results, conclusions can be drawn about different strain and doping values: CVD grown MoS$_{2}$ appears to be more strained and less n-doped compared to exfoliated MoS$_{2}$. However, the actual origin for the apparent differences in PL and Raman is the presence/absence of an intercalated water film. As soon as a water film is present under the CVD grown MoS$_{2}$, the difference in charge carrier density vanishes. Both, the differences in PL intensity and strain/defect density also become much smaller then. The only difference we could identify is a higher/lower density of non-doping defects in the CVD-grown material. Therefore, we draw the clear conclusion from our data that the \textit{intrinsic} properties of MoS$_{2}$ grown by CVD and prepared by exfoliation, respectively, are only marginal. 

Our data shows that CVD processes run with MoO$_{3}$ as precursor typically result in samples with a much larger variation with respect to the strain/defect density values. This indicates the superiority of the AHM precursor for the growth of MoS$_{2}$ via CVD and leads us to believe that presumably many other 2D TMDCs may be grown with a similarly high quality if the right precursor can be found. 

We could show, that MoS$_{2}$ on graphite is indeed a good model system for the investigation of MoS$_{2}$-graphene heterostructures. In this case, CVD grown MoS$_{2}$ interacts much stronger with the substrate. A decoupling by an intercalated water film, as typically occurs after exfoliation or transfer, is not sufficient to ensure that the MoS$_{2}$ is no longer influenced by the substrate. These findings render heterostructures directly grown via CVD even more promising for energy conversion, sensing, and spintronic devices as one would have expected from proof-of-principle experiments with stacked/transferred van der Waals materials, of which at least one is often exfoliated \cite{Dankert.2017, Roy.2013, Yu.2013, Cho.2015, Zhao.2018}.

\section{Methods}
\subsection{Sample preparation}

The single layer MoS$_{2}$ sheets are either prepared by standard exfoliation technique (scotch tape method) \cite{Novoselov.2004, Novoselov.2005} (blue squares in the figures) from a natural bulk MoS$_{2}$ crystal or by CVD in a three-zone split tube furnace (ThermConcept ROK 70/750/12-3z). Two processes were used: (i) molybdenum trioxide (MoO$_{3}$) powder as molybdenum source and referred to as MoO$_{3}$ prcoess in the following (green symbols), and (ii) ammonium heptamolybdate (AHM) as precursor for the molybdenum source in the AHM process (light blue symbols). The used recipes follow:

\paragraph{MoO$_{3}$ Process}
This process is a variation of the classical CVD method developed by Lee et al.\cite{Lee.2012}. Two heating zones were used here. A ceramic boat with 50~mg sulfur powder (Sigma Aldrich, 99.98 \%) is positioned in the upstream heating zone, a ceramic boat with <1~mg MoO$_{3}$ powder (Alfa Aesar, 99.95\%) and SiO$_{2}$ susbtrates, which are first cleaned in an ultrasonic bath in ethanol and then treated with perylene-3,4,9,10-tetracarboxylic acid tetrapotassium acid salt (PTAS, 2D semiconductors) as the seeding promoter, is positioned in the adjacent downstream heating zone. After an ambient pressure argon flushing of the tube for 30 minutes at an Ar  flow of 50~Ncm$^{3}$/min, which is maintained for the whole process, the heating zone with the MoO$_{3}$ powder and the substrates is heated up to a maximum temperature of 750-800~\textdegree C with a rate of 1600~\textdegree C/h. With a delay of 30~min the sulfur heating zone is heated to 180~\textdegree C within 5~min. After a holding time of 25~min at the maximum temperature, the furnace is opened for rapid cooling.

\paragraph{AHM Process}
With this process type according to Han et al. \cite{Han.2015}, the molybdenum source, also in the form of MoO$_{3}$, is provided in a different way than in the process described above. Instead of adding MoO$_{3}$ directly into the process system in powder form, it is first produced from water-soluble ammonium heptamolybdate (AHM, Sigma Aldrich) in an additional decomposition step \cite{Hanafi.1981, Wienold.2003}. Much less than 1~\textmu L (<< 0.2~mg AHM) is dropleted from a 50:50 mixture of a saturated AHM solution and deionized water onto the previously cleaned substrates and then heated for 30 min at 300~\textdegree C to convert AHM to MoO$_{3}$. In this way, reproducible and smaller quantities of molybdenum can be introduced into the CVD process system. The seeding promoter cholic acid sodium salt (Sigma Aldrich) is then spun on. After positioning the substrate with the molybdenum source in the downstream heating zone and 50~mg sulfur powder in the upstream heating zone the process can be operated as above. However, for the process slightly different optimal parameters were found: Ar flow is hold at 500~Ncm$^{3}$/min for the whole process (including at least 15 min flushing time, the process and the cooling time). Both heating zones are heated up within 11 minutes to the process temperatures of 150~\textdegree C and 750~\textdegree C for the heating zone containing the sulfur source and the heating zone containing the the molybdenum source and the substrates respectively. These process temperatures are held for 19 minutes before the furnace is opened for rapid cooling.

\paragraph{Transfer}
In order to prepare some of the samples, different transfer techniques are used for the 2D materials MoS$_{2}$ and graphene (Graphenea). To transfer MoS$_{2}$, a poly(methyl methacrylate) (PMMA) based wet transfer was used. In detail, a thin PMMA layer (ARP 671.05, Allresist GmbH) was spun on the MoS$_{2}$ samples and annealed (5~min, 100~\textdegree C). In order to remove the PMMA/MoS$_{2}$ the SiO$_{2}$ is etched by a KOH solution (0.7 mol/L). After replacing the etching solution with deionized water, the floating PMMA/MoS$_{2}$ stack is scooped out by another clean SiO$_{2}$ substrate. The PMMA is removed by an acetone bath. For transferring graphene a polymer free wet transfer is used, which is reported in our previous work \cite{Madau.2019}. In case of growing MoS$_{2}$ onto the graphene-SiO$_{2}$ samples, the samples were pre-annealed in activated carbon for 20~min at 100~\textdegree C.

\subsection{Characterization}

Confocal Raman and PL spectroscopy were performed with a Raman microscope (Renishaw InVia) with a laser wavelength of 532~nm and a spot size of 1~\textmu m. Power densities of the laser were typically $\sim$0.06~mW/\textmu m$^{2}$, but had to be reduced by several orders of magnitude for samples with extreme PL intensity to avoid saturation of the detector. Atomic force microscope (AFM) measurements were performed on a Veeco Dimension 3100 AFM in Tapping Mode using Nanosensors PPP-NCHR tips and on a Bruker Dimension Icon in PeakForce Tapping Mode with Bruker ScanAsyst-Air tips. The latter mode is based on the recording of a large number of force-distance curves, thus allowing the simultaneous spatial resolution of mechanical properties (deformation, adhesion, ...) in addition to topography.


\begin{acknowledgement}

The authors acknowledge support from the German Research Foundation (DFG) by funding SCHL 384/20-1 (project number 406129719). PeakForce AFM, Raman and PL spectroscopy was performed at the Interdisciplinary Center for Analytics on the Nanoscale (ICAN), a core facility funded by the German Research Foundation (DFG, reference RI\_00313).

\end{acknowledgement}

\begin{suppinfo}

The following files are available free of charge.
\begin{itemize}
  \item Supplementary Material - MoS$_{2}$ CVD vs Exfoliation: More AFM images for exemplary MoS$_{2}$ thickness and substrate roughness evaluation, additional PeakForce AFM channel
\end{itemize}

\end{suppinfo}


\bibliography{Bib}

\newpage
\section*{TOC Graphic}
\includegraphics[scale=1.0]{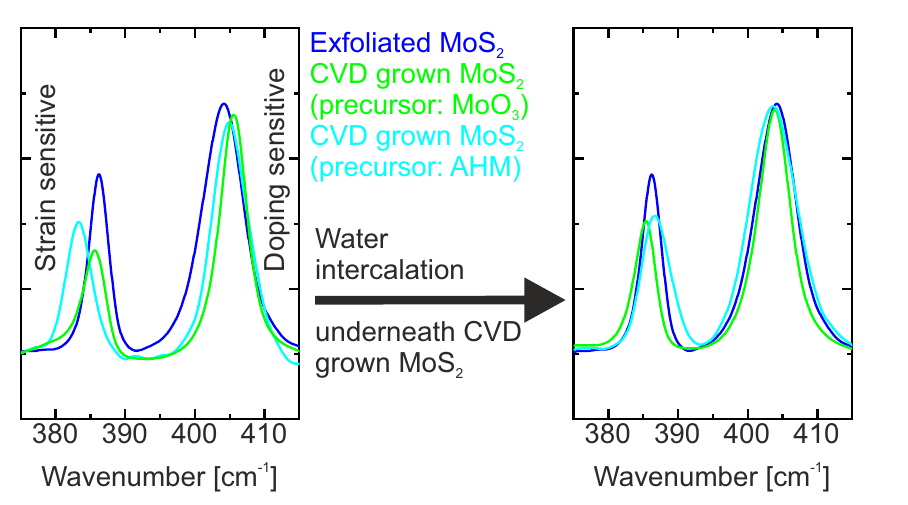}

\end{document}